\documentclass[12pt]{article}
\def \beq{\begin{equation}}
\def \eeq{\end{equation}}
\def \ew{reprinted in {\it The Eightfold Way, op.\ cit}}
\def \ite{{\it et al.}}
\def \efi{Enrico Fermi Institute Report No.~}

\begin{document}
\baselineskip 20pt
\centerline{\bf THE EIGHTFOLD WAY
\footnote{\efi 01-41, hep-ph/0109241, to be published in {\it Macmillan
Encylopedia of Physics, Supplement:  Elementary Particle Physics},
edited by John S. Rigden, Jonathan Bagger, and Roger H. Stuewer
(Macmillan Reference USA, New York, 2002).}}
\centerline{Jonathan L. Rosner}
\medskip

The Eightfold Way is the name coined by Murray Gell-Mann (1961) to describe a
classification scheme of the elementary particles devised by him and
Yuval Ne'eman (1961).  The name, adopted from the Eightfold Path of Buddhism,
refers to the eight-member families to which many sets of particle belong.

In the 1950s Gell-Mann and Kazuo Nishijima invented a scheme to explain a
``strange'' feature of certain particles; they appeared to be easily produced in
cosmic-ray and accelerator reactions, but decayed slowly, as if something were
hindering their decays.  These particles were assumed to carry a property
known as {\it strangeness} which would be preserved in production but could
be changed in decays.  Two examples of plots of electric charge (in units of
the fundamental charge $|e|$) versus strangeness for particles known in the
late 1950s are the following:

\renewcommand{\arraystretch}{1.4}
\begin{center}
\begin{tabular}{|c c c c|c|c c c c|} \cline{1-4} \cline{6-9}
\multicolumn{4}{|c|}{Mesons:} & & \multicolumn{4}{c|}{Baryons:} \\
Strangeness: & \multicolumn{3}{c|}{Particle:} & &
Strangeness: & \multicolumn{3}{c|}{Particle:} \\ \cline{1-4} \cline{6-9}
1  &         & K$^0$   &  K$^+$ &  & 0 & & n & p \\
0  & $\pi^-$ & $\pi^0$ & $\pi^+$ & & $-1$~~ & $\Sigma^-$ & $\Sigma^0,\Lambda$ &
 $\Sigma^+$ \\
$-1$~~ & K$^-$ & $\bar{\rm K}^0$ & & & $-2$~~ & $\Xi^-$ & $\Xi^0$ & \\
\cline{1-4} \cline{6-9}
Charge: & $-1$~~ & 0 & 1 & & Charge: & $-1$~~ & 0 & 1 \\ \cline{1-4}
\cline{6-9}
\end{tabular}
\end{center}

{\it Mesons} include the $\pi$ particles, known as {\it pions}, whose existence
was proposed by Hideki Yukawa in 1935 to explain the strong nuclear force, and
the K particles (also known as kaons), discovered in cosmic radiation in the
1940s. Pions and kaons weigh about one-seventh and one-half as much as protons,
respectively.  {\it Baryons} (the prefix {\it bary-} is Greek for {\it heavy})
include the proton p, the neutron n, and heavier
relatives $\Lambda$ (lambda), $\Sigma$ (sigma), and $\Xi$ (xi), collectively
known as {\it hyperons} and discovered in the 1940s and 1950s.  The rationale
for these families was sought through {\it symmetries of the strong
interactions}.

According to the Gell-Mann--Nishijima scheme, reactions in which these
particles are produced must have equal total strangeness on each side.  For
example, a $K^0$ and a $\Lambda$ can be produced by the reaction
$$
\pi^-~({\rm S} = 0) + {\rm p}~({\rm S} = 0) \to {\rm K}^0~({\rm S} = 1)
+ \Lambda~({\rm S} = -1)~~~.
$$
This scheme thus explained another curious feature of the ``strange particles'':
They never appeared to be produced singly in reactions
caused by protons, neutrons, and $\pi$ mesons.

In the 1930s, Werner Heisenberg and others had recognized that the similarities
of the proton and neutron with respect to their nuclear interactions and masses
could be described by a quantity known as {\it isotopic spin}.  This quantity,
called {\it isospin} for short, is analogous to ordinary spin with the proton's
isospin pointing ``up'' and the neutron's pointing ``down.''  Mathematically,
isospin is described by a {\it symmetry group}, i.e., a set of transformations
which leaves interactions unchanged, known as SU(2).  The ``2'' refer to the
the proton and neutron.

Families whose members are related to one another by SU(2) transformations
can have any number of members, including the two-member family to which the
proton and neutron belong.  Collectively, p and n are known as {\it nucleons},
and denoted by the symbol N.  Isospin predicts that certain sets of particles
with different charges (e.g., K or $\Sigma$) should have similar masses and
strong interactions, as is observed.

Shoichi Sakata (1956) proposed that mesons were composed of the proton p, the
neutron n, the lambda $\Lambda$, and corresponding antiparticles, with binding
forces so large as to overcome most of their masses.
Thus, for example, the K$^+$ would be p$\overline{\Lambda}$.
(The bar over a symbol denotes its antiparticle; electric charges and
strangeness of antiparticles are opposite to those of the corresponding
particles.)  The remaining known baryons (the $\Sigma$ and $\Xi$) had to be
accounted for in more complicated ways.  The Sakata model had the symmetry
known as SU(3), where ``3'' referred to p, n, and $\Lambda$.

Gell-Mann and Ne'eman recognized that if electric charge were to be part of the
SU(3) description, particles whose electric charges were integer multiples of
$|e|$ could belong only to certain families.  The simplest of these contained
one, eight, and ten members.  Other families, such as those containing three
and six members, would have fractionally-charged members, and fractional
charges had never been seen in nature.  Both the mesons and the baryons
mentioned above would then have to belong to eight-member families.  The
baryons fit such a family exactly, leading Gell-Mann to call his scheme the
``Eightfold Way.''  In addition to the seven mesons shown, there would have to
be an eighth meson, neutral and with zero strangeness.  This particle,
now called the $\eta$ (eta), was discovered in 1961.  

A consequence of the Eightfold Way for describing mesons and baryons was that
their masses M could be related to one another by formulae proposed by
Gell-Mann and by Susumu Okubo (1962):
$$
{\rm Mesons:}~~~~4 {\rm M(K) = M}(\pi) + 3 {\rm M}(\eta)~~~;
$$
$$
{\rm Baryons:}~~~~2[{\rm M(N) + M}(\Xi)] = {\rm M}(\Sigma) + 3 {\rm M}
(\Lambda)~~~.
$$
These formulae, particularly the one for baryons, were obeyed quite well.
More evidence for SU(3) soon arrived from another experimental discovery.

Certain baryons known as $\Delta$ (delta), $\Sigma^*$ (sigma-star), and $\Xi^*$
(xi-star) appeared to fit into a ten-member family, which would be completed
by a not-yet-observed particle known as the $\Omega^-$ (omega-minus):

\begin{center}
\begin{tabular}{|c c c c c|} \hline
Strangeness: & \multicolumn{4}{c|}{Baryon:} \\ \hline
0 & $\Delta^-$ & $\Delta^0$ & $\Delta^+$ & $\Delta^{++}$ \\
$-1$~~ & $\Sigma^{*-}$ & $\Sigma^{*0}$ & $\Sigma^{*+}$ & \\
$-2$~~ & $\Xi^{*-}$ & $\Xi^{*0}$ & & \\
$-3$~~ & ($\Omega^-$) & & & \\ \hline
Charge: & $-1$~~ & 0 & 1 & 2 \\ \hline
\end{tabular}
\end{center}

The mass of the $\Omega^-$
could be anticipated within a few percent because the Gell-Mann--Okubo mass
formula for these particles predicted
$$
M(\Omega^-) - M(\Xi^*) = M(\Xi^*) - M(\Sigma^*) = M(\Sigma^*) - M(\Delta)~~.
$$ 
An experiment at Brookhaven National Laboratory (Barnes \ite, 1964)
detected this particle with the predicted mass through a decay that left no
doubts as to its nature.

An early application of the Eightfold Way, building upon suggestions by
Gell-Mann and Maurice L\'evy (1960) and by Gell-Mann (1962), was made by
Nicola Cabibbo (1963) to certain decays of baryons, which showed that SU(3)
symmetry could be used to describe not only the existence and masses of
particles but also their interactions.

Underlying the success of the Eightfold Way and the symmetry group SU(3) is
the existence of fundamental subunits of matter, called {\it quarks} by
Gell-Mann (1964) and {\it aces} by their co-inventor, George Zweig (1964).
These objects can belong to a family of three {\it fractionally-charged}
members u (``up''), d (``down''), and s (``strange''):

\begin{center}
\begin{tabular}{|c c c|} \hline
Strangeness: & \multicolumn{2}{c|}{Quark:} \\ \hline
 0 & d & u \\
$-1$~~ & s & \\ \hline
Charge: & $-1/3$~~ & 2/3 \\ \hline
\end{tabular}
\end{center}

The fact that fractionally-charged objects have not been seen in nature
requires quarks to combine with one another in such as way as to produce only
integrally-charged particles.  This is one successful prediction of the theory
of the strong interactions, {\it quantum chromodynamics} (QCD).  Baryons are
made of three quarks, while mesons are made of a quark and an {\it antiquark}
(with reversed charge and strangeness).  For example, the $\Delta^{++}$ is
made of uuu; the $\Delta^-$ is made of ddd; the $\Omega^-$ is made of sss;
and the $K^+$ is made of u$\bar{\rm s}$.

Other quarks --- {\it charm} (c), {\it bottom} (b), and {\it top} (t) ---
were discovered subsequently.  They are much heavier than u, d, and s.  The
approximate SU(3) symmetry of particles containing u, d, and s quarks
remains a useful guide to properties of the strong interactions.
\newpage

\centerline{\bf BIBLIOGRAPHY}
\baselineskip 18pt

\leftline{\underline{Book}:}

\noindent
GELL-MANN, M. and NE'EMAN, Y. {\it The Eightfold Way,} New York:  W. A.
Benjamin, 1964.
\bigskip

\noindent
\leftline{\underline{Journal articles}:}

\noindent
BARNES, V. E. {\it et al.} ``Observation of a Hyperon With Strangeness Minus
Three.'' {\it Phys.\ Rev.\ Letters} 12, 204--206 (1964), \ew, pp.\ 88-92.
\medskip

\noindent
CABIBBO, N. ``Unitary Symmetry and Leptonic Decays.'' {\it Phys.\ Rev.\
Letters} 10, 531-533 (1963), \ew, pp.\ 207--209.
\medskip

\noindent
GELL-MANN, M. ``The Eightfold Way:  A Theory of Strong Interaction Symmetry.''
California Institute of Technology Report CTSL-20, 1961 (unpublished),
\ew, pp.\ 11--57.
\medskip

\noindent
GELL-MANN, M. ``Symmetries of Baryons and Mesons.'' {\it Phys.\ Rev.} 125,
1067--1084 (1962), \ew, pp.\ 216--233.
\medskip

\noindent
GELL-MANN, M. ``A Schematic Model of Baryons and Mesons.'' {\it Phys.\
Letters} 8, 214--215 (1964), \ew, pp.\ 168--169.
\medskip

\noindent
GELL-MANN, M. and L\'EVY, M. ``The Axial Vector Current in Beta Decay.''
{\it Nuovo Cimento} 16, 705 (1960).
\medskip

\noindent
NE'EMAN, Y.  ``Derivation of Strong Interactions from a Gauge Invariance.''
{\it Nucl.\ Phys.} 26, 222-229 (1961), \ew, pp.\ 58--65.
\medskip

\noindent
OKUBO, S. ``Note on Unitary Symmetry in Strong Interactions.'' {\it Prog.\
Theor.\ Phys.} (Kyoto) 27, 949--966 (1962), \ew, pp.\ 66-83.
\medskip

\noindent
SAKATA, S. ``On a Composite Model for the New Particles.''
{\it Prog.\ Theor.\ Phys.} (Kyoto) 16, 686--688 (1956).
\medskip

\noindent
ZWEIG, G.  CERN Reports 8182/TH 401 and 8419/TH 412 (unpublished), second paper
reprinted in {\it Developments in the Quark Theory of Hadrons}, edited by
D. B. Lichtenberg and S. P. Rosen, Nonantum, Mass: Hadronic Press, 1980,
v.\ 1, pp. 22--101.
\end{document}